\documentclass[runningheads]{llncs}
\usepackage{graphicx}
\usepackage[implicit=false]{hyperref}
\begin{document}
\title{Selection of Optimal Parameters in the Fast K-Word Proximity Search Based on Multi-component Key Indexes}
\titlerunning{Fast K-Word Proximity Search Based on Multi-component Key Indexes}
%
\author{Alexander B. Veretennikov\inst{1}\orcidID{0000-0002-3399-1889} }
\authorrunning{A. B. Veretennikov}
%
\institute{Ural Federal University, 620002 Mira street, Yekaterinburg, Russia
\email{alexander@veretennikov.ru}
}
\maketitle              

This is a pre-print of a contribution published in 
Supplementary Proceedings of the XXII International Conference on Data Analytics and Management in Data Intensive Domains (DAMDID/RCDL 2020), Voronezh, Russia, October 13--16, 2020, P. 336--350,
published by CEUR Workshop Proceedings. The final authenticated version is available online at: 
\href{http://ceur-ws.org/Vol-2790/}{http://ceur-ws.org/Vol-2790/}.

\noindent
Indexing: Scopus.

\noindent
See also: \href{http://www.veretennikov.org}{http://www.veretennikov.org}.

\begin{abstract}
Proximity full-text search is commonly implemented in contemporary full-text search systems. Let us assume that the search query is a list of words. It is natural to consider a document as relevant if the queried words are near each other in the document. The proximity factor is even more significant for the case where the query consists of frequently occurring words. Proximity full-text search requires the storage of information for every occurrence in documents of every word that the user can search. For every occurrence of every word in a document, we employ additional indexes to store information about nearby words, that is, the words that occur in the document at distances from the given word of less than or equal to the $MaxDistance$ parameter. We showed in previous works that these indexes can be used to improve the average query execution time by up to 130 times for queries that consist of words occurring with high-frequency. In this paper, we consider how both the search performance and the search quality depend on the value of $MaxDistance$ and other parameters. Well-known GOV2 text collection is used in the experiments for reproducibility of the results. We propose a new index schema after the analysis of the results of the experiments.

\end{abstract}

\section{Introduction}
In full-text search, a query is a list of words. The result of the search is a list of documents containing these words. Consider a query that consists of words occurring with high-frequency. In \cite{Veretennikov:DAMDID:2018}, we improved the average query processing time by a factor of 130 for these queries relative to traditional inverted indexes. A methodology for high-performance proximity full-text searches that covers different types of queries was discussed in \cite{Veretennikov:IntelliSys:2018}.

The factor of proximity or nearness between the queried words in the indexed document plays an important role in modern information retrieval \cite{Yan:2010:ETP:1871437.1871593,Lu:2016:EEH:2970398.2970404,10.1145/2094072.2094077}. We assume that a document should contain query words near each other to be relevant for the user in the context of the search query. Taking this factor into account is essential if the query consists of frequently occurring words.

\begin{figure}[h]
 \setlength{\abovecaptionskip}{1pt}
 \setlength{\belowcaptionskip}{-10pt}
  \centering
  \includegraphics[width=0.6\linewidth]{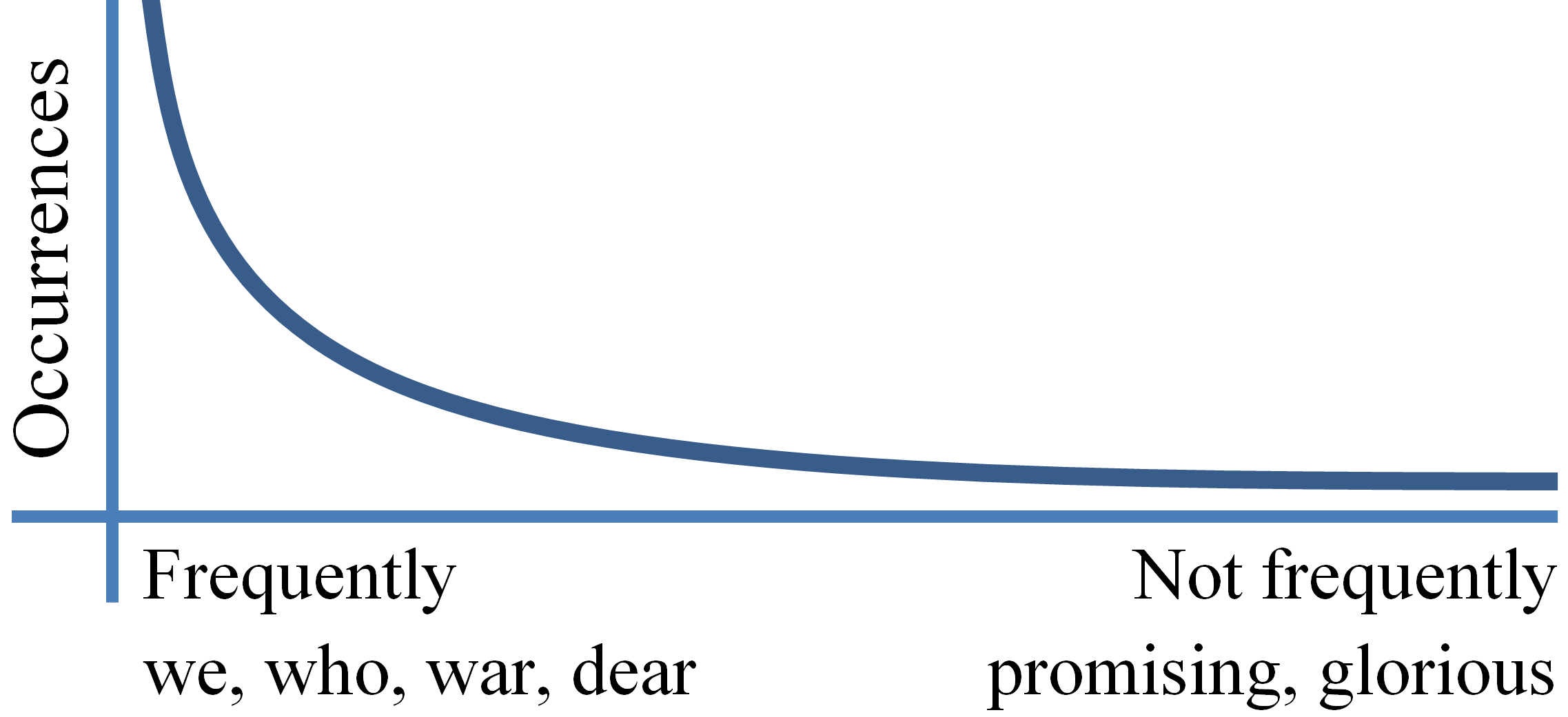}
  \caption{Example of a word frequency distribution.}
  \label{VeretennikovA-image-Zipf}
\end{figure}

Some words occur in texts significantly more frequently than others. We can illustrate this \cite{Veretennikov:IntelliSys:2018} by referring to Zipf's law \cite{Zipf:1932}. An example of a typical word occurrence distribution is presented in Fig.~\ref{VeretennikovA-image-Zipf}. The horizontal axis represents different words in decreasing order of their occurrence in texts. On the vertical axis, we plot the number of occurrences of each word. This peculiarity of language has a strong influence on the search performance.

The inverted index \cite{Zobel:2006:IFT:1132956.1132959,BorodinGIN} is a commonly used data structure for full-text searches. The traditional inverted index contains records $(ID, P)$, where $ID$ is the identifier of the document and $P$ is the position of the word in the document, for example, an ordinal number of the word in the document. This record corresponds to an occurrence of a word in a document. These $(ID, P)$ records are called ``postings''. The inverted index enables us to obtain a list of postings that corresponds to the given word. These lists of postings are used for the search.

For proximity full-text searches, we need to store the $(ID, P)$ record for every occurrence of every word in the indexed document \cite{Sadakane:2001,KWordProximityEncrypted:2016,UtilizingDistinctTerms:2014}. In other words, for proximity searches, we require a word-level inverted index instead of a document-level index \cite{KWordProximityEncrypted:2016}. Consequently, if a word occurs frequently in the texts, then its list of postings is long \cite{Veretennikov:IntelliSys:2018}. The query search time is proportional to the number of occurrences of the queried words in the indexed documents. To process a search query that contains words occurring with high-frequency, a search system requires much more time, as shown on the left side of Fig.~\ref{VeretennikovA-image-Zipf}, than a query that contains only ordinary words, as shown on the right side of Fig.~\ref{VeretennikovA-image-Zipf}.
  
According to \cite{10.1145/1476589.1476628}, we can consider a full-text query as a ``simple inquiry'' \cite{Veretennikov:IntelliSys:2018}. In this case, we may require that the search results be provided within two seconds, as stated in \cite{10.1145/1476589.1476628}, to prevent the interruption of the thought continuity of the user.
To enhance the performance, the following approaches can be used.

Early-termination approaches can be employed for full-text searches \cite{Anh:2001:VRE:383952.383957,10.1145/2808194.2809477}. However, these methods are not effective in the case of proximity full-text searches \cite{Veretennikov:DAMDID:2018}. Usually, early-termination approaches are applied to document-level indexes. It is difficult to combine the early-termination approach with the incorporation of term-proximity data into relevance models.

Additional indexes can improve the search performance. In \cite{10.1145/2766462.2767769,Williams:2004:FPQ:1028099.1028102}, additional indexes were used to improve phrase searches. However, the approaches reported in \cite{10.1145/2766462.2767769,Williams:2004:FPQ:1028099.1028102} cannot be used for proximity full-text searches. Their area of application is limited by phrase searches. We have overcome this limitation. 

With our additional indexes, an arbitrary query can be executed very fast and quickly \cite{Veretennikov:IntelliSys:2018}.

In an example given in \cite{Veretennikov:DAMDID:2018}, we indexed a subset of the Project Gutenberg web site using Apache Lucene and Apache Tika and performed some searches. A query that contained words occurring with high-frequency was evaluated within 21 sec. On the other hand, another query that contained ordinary words was evaluated within 172 milliseconds. This difference is considerable. However, our algorithms reported in \cite{Veretennikov:DAMDID:2018,Veretennikov:IntelliSys:2018} can help address this issue.

The goals and questions of this paper are as follows:

\vspace{-\topsep}
\vspace{1pt}
\begin{enumerate}
    \setlength{\itemsep}{1pt}
    \setlength{\parskip}{1pt}
    \setlength{\parsep}{1pt}
\item[1)]	We need to examine how search performance is improved with consideration of different values of $MaxDistance$ and other parameters.
\item[2)]	We need to investigate search performance on the commonly used text collection.
\item[3)]	Does the search performance depend on the document size?
\item[4)]	What are other factors that can affect the performance?
\item[5)]	We evaluate the performance with respect to both short and long queries.
\end{enumerate}

Key points of our research are the following.
We use the word-level index.
We use the DAAT approach \cite{10.1016/j.ipm.2016.03.005,Jiang:2015:TEI:2839534.2840112}.
We include in the indexes information about all lemmas.
Our indexes support incremental updates.
We can store one posting list in several data streams.
We read entire posting lists when searching and no early-termination are used.

\section{Lemmatization and Lemma Type}
For this paper, we use an English dictionary with approximately 92 thousand English lemmas. This dictionary is used by a morphology analyzer. The analyzer produces a list of numbers of lemmas, that is, basic or canonical forms, for every word from the dictionary. Usually, a word has one lemma, but some words have several lemmas. For example, the word ``mine'' has two lemmas, namely, ``mine'' and ``my''.

Consider an array of all lemmas. Let us sort all lemmas in decreasing order of their occurrence frequency in the texts. We call the result of sorting the $FL$-list \cite{Veretennikov:IntelliSys:2018}. The number of a lemma $w$ in the $FL$-list is called its $FL$-number \cite{Veretennikov:IntelliSys:2018} and is denoted by $FL(w)$. Let us say that lemma ``earth'' $>$ ``day'' because $FL(\textnormal{earth}) = 309$, $FL(\textnormal{day}) = 199$, and $309 > 199$. We use the $FL$-numbers to define the order of the lemmas in the collection of all lemmas.

In our search methodology \cite{Veretennikov:DAMDID:2018}, we defined three types of lemmas, namely, stop lemmas, frequently used lemmas and ordinary lemmas.

The first $SWCount$ most frequently occurring lemmas are stop lemmas. Examples include ``earth'', ``yes'', ``who'', ``day'', ``war'', ``time'', ``man'' and ``be''.

The second $FUCount$ most frequently occurring lemmas are frequently used lemmas. Examples include ``red'', ``beautiful'', and ``mountain''.

All other lemmas are ordinary lemmas, e.g., ``fiber'' and ``undersea''.

$SWCount$ and $FUCount$ are the parameters. 
We use $SWCount = 500$ and $FUCount = 1050$ in the experiments described here.
Let $MaxDistance$ be a parameter that can take a value of 5, 7, 9 or even greater.

The value of $SWCount$ is very near 421 from \cite{10.1145/378881.378888}. However, as we include information about all lemmas in the indexes, we can use very different values of the parameters.
If an ordinary lemma, $q$, occurs in the text so rarely that $FL(q)$ is irrelevant, then we can say that $FL(q) = \sim$. Here, ``$\sim$'' denotes a large number.

\section{Can stop words be skipped?}
Let us discuss the stop-word approach, in which some words occurring with high-frequency or their lemmas are excluded from the search. We do not agree with this approach. A word cannot be excluded from the search because even a word occurring with high-frequency can have a specific meaning in the context of a specific query \cite{Veretennikov:DAMDID:2018,Williams:2004:FPQ:1028099.1028102}. Therefore, excluding some words from the search can lead to search quality degradation or unpredictable effects \cite{Williams:2004:FPQ:1028099.1028102}. Additionally, stop words are often employed in higher-order term proximity feature models \cite{10.1145/3018661.3018726}. 

Consider the query ``who are you who'' \cite{Veretennikov:IntelliSys:2018}. The Who are an English rock band, and ``Who are You'' is one of their works. The word ``Who'' has a specific meaning in the context of this query. Therefore, in our approach, we include information about all of the words in the indexes. Moreover, we can easily see that modern search systems, such as Google, do not skip stop words in a search.

\section{The types of additional indexes}
The three-component key $(f, s, t)$ index \cite{Veretennikov:IntelliSys:2018} is the list of the occurrences of the lemma $f$ for which lemmas $s$ and $t$ both occur in the text at distances that are less than or equal to the $MaxDistance$ from $f$. Each posting includes the distance between $f$  and $s$ in the text and the distance between $f$ and $t$ in the text. An $(f, s, t)$ index is created only for the case in which $f \leq s \leq t$. Here, $f$, $s$, and $t$ are all stop lemmas.

In \cite{Veretennikov:DAMDID:2018}, we considered queries consisting of high-frequency occurring words. In addition, we showed that the average query execution time can be improved with three-component key indexes by up to 15.6 times relative to the time necessary using two-component key indexes only. Therefore, $(f, s, t)$ indexes are required if we need to search queries that contain high-frequently occurring words.

The two-component key $(w, v)$ index \cite{Veretennikov:IntelliSys:2018} is the list of occurrences of the lemma $w$ for which lemma $v$ occurs in the text at a distance that is less than or equal to the $MaxDistance$ from $w$. Each posting includes the distance between $w$ and $v$ in the text. Here, $w$ denotes a frequently used lemma, and $v$ denotes a frequently used or ordinary lemma.

Let us consider the traditional index with near stop word (NSW) records. For each occurrence of each ordinary or frequently used lemma in each document, we include a posting record $(ID, P, \textnormal{NSW})$ in the index. $ID$ can be the ordinal number of the specific document, and $P$ is the position of the word in the document, e.g., the ordinal number of the word in the document. The NSW record contains information about all high-frequency lemmas, that is, stop lemmas, occurring near position $P$ in the document (at a distance that is less than or equal to the $MaxDistance$ from $P$).
Examples of the indexes are given in \cite{Veretennikov:IntelliSys:2018}.

The posting list of a key is stored in several data streams \cite{Veretennikov:IntelliSys:2018}. For the traditional index with NSW records, we can use up to three streams for one key: one stream for $(ID)$, one for $(P)$ and one for (NSW). On the other hand, for this index, we can use two streams: the first is $(ID, P)$, and the second is (NSW). The actual choice depends on the length of the posting list. For short lists, we use two streams; for long lists, we use three streams. This architecture allows us to skip NSW records when they are not required. For the $(w, v)$ and $(f, s, t)$ indexes, we use one or two data streams for every key.

\section{Experiment}

\subsection{Indexes, collection and environment}
We create the following indexes:

$Idx0$: the traditional inverted index without any enhancements, such as NSW records. The total size is 143 GB. This value includes the total size of indexed texts in compressed form, which is 57.3 GB.

$Idx5$: our indexes, including the traditional inverted index with the NSW records and the $(w, v)$ and $(f, s, t)$ indexes, where $MaxDistance$ = 5. The total size is 1.29 TB, the total size of the $(w, v)$ index is 104 GB, the total size of the $(f, s, t)$ index is 727 GB, and the total size of the traditional index with NSW records is 192 GB.

$Idx7$: our indexes, where $MaxDistance$ = 7. The total size is 2.16 TB, the total size of the $(w, v)$ index is 148 GB, the total size of the $(f, s, t)$ index is 1.422 TB, and the total size of the traditional index with NSW records is 239 GB.

$Idx9$: our indexes, where $MaxDistance$ = 9. The total size is 3.27 TB, the total size of the $(w, v)$ index is 191 GB, the total size of the $(f, s, t)$ index is 2.349 TB, and the total size of the traditional index with NSW records is 283 GB.

For the experiment, GOV2 \cite{TREC:2006} text collection and the following queries are used: 
title queries from TREC Robust Task 2004 (with 250 queries in total),
title queries from TREC Terabyte Task from 2004 to 2006 (with 150 queries in total),
title queries from TREC Web Task from 2009 to 2014 (with 300 queries in total),
queries from TREC 2007 Million Query Track (10000 queries in total).

The total size of the query set after duplicate removal is 10 665 queries.
GOV2 text collection contains 25 million documents. The total size of the collection is approximately 426 GB, and after HTML tag removal, there is approximately 167 GB of plain text. The average document text size is approximately 7 KB.

We used the following computational resources:
CPU: Intel(R) Core(TM) i7 CPU 920 @ 2.67 GHz.
HDD: 7200 RPM. RAM: 24 GB.
OS: Microsoft Windows 2008 R2 Enterprise.

The query set can be divided into the following subsets depending on lemmas in a concrete query \cite{Veretennikov:IntelliSys:2018}.
All queries are evaluated within one program thread. 

If we have a query with a length greater than $MaxDistance$, then we should divide it into several parts. For example, when the value of $MaxDistance$ is 5, then the query ``to be or not to be that is the question'' should be rewritten as ``(to be or not to) (be that is the question)'', and these two parts should be evaluated independently; then, the results should be combined. 

\subsection{Q1. Only stop lemmas}
Every query in the subset contains only stop lemmas. There are 119 queries in this subset.
Examples include the following:
to be or not to be that is the question,
kids earth day activities.
With $FL$-numbers, we have the following queries:
[to: 9] [be: 7] [or: 38] [not: 64] [to: 9] [be: 7] [that: 40] [be: 7] [the: 1] [question: 305]
and 
[kid: 447] [earth: 309] [day: 199] [activity: 247].

For these queries, the $(f, s, t)$ indexes are used \cite{Veretennikov:DAMDID:2018}. 

\begin{figure}[h]
  \setlength{\abovecaptionskip}{1pt}
  \setlength{\belowcaptionskip}{-10pt}
  \centering
  \includegraphics[width=\linewidth]{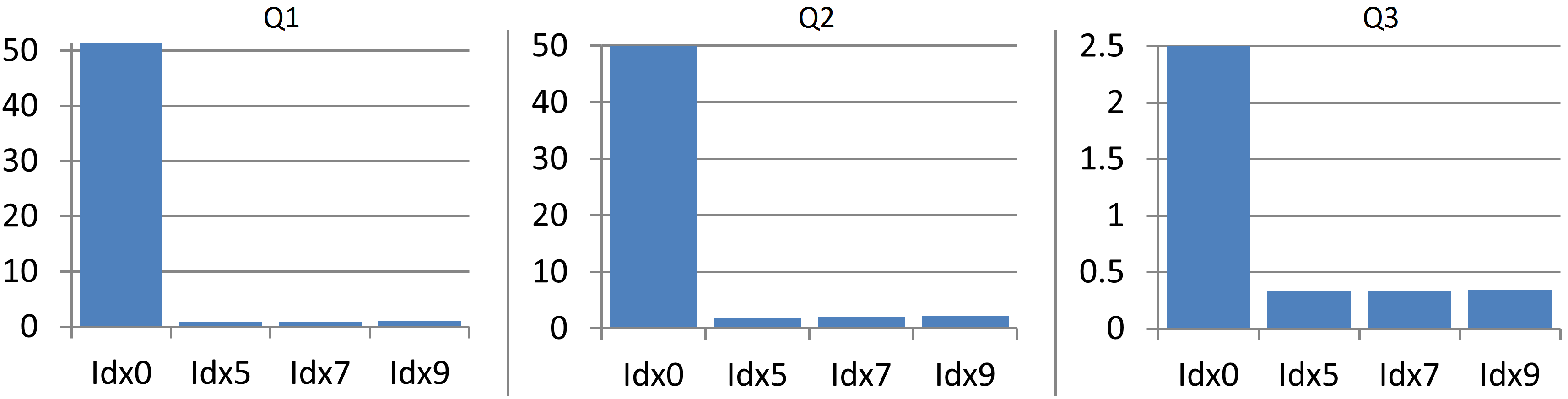}
  \caption{The average query execution times for $Idx0$, $Idx5$, $Idx7$, and $Idx9$ (seconds); the query subsets Q1, Q2, Q3.}  
  \label{VeretennikovA-image-step-Search-Q1Q2Q3}
\end{figure}

Average query processing times:

$Idx0$: 51.4 s, $Idx5$: 0.82 s, $Idx7$: 0.86 s, $Idx9$: 1.05 s (see Fig.~\ref{VeretennikovA-image-step-Search-Q1Q2Q3}).

Average data read sizes per query:

$Idx0$: 1.3 GB, $Idx5$: 11.1 MB, $Idx7$: 15.6 MB, $Idx9$: 20.1 MB.

Average numbers of postings per query:

$Idx0$: 317.8 million, $Idx5$: 0.88 million, $Idx7$: 1.15 million, $Idx9$: 1.5 million.

We improved the query time by a factor of 62.7 with $Idx5$, by a factor of 59.4 with $Idx7$, and by a factor of 48.7 with $Idx9$ in comparison with $Idx0$.
  
\subsection{Q2. Stop and frequently used and/or ordinary lemmas}

Every query in the subset contains one or several stop lemmas. The query also contains some other lemmas that may be frequently used or ordinary. There are 7 244 queries in the subset.
Examples include the following:
History of Physicians in America.
With $FL$-numbers, we have the following query:

[history: 598] [of: 4] [physician: 1760] [in: 14] [America: 1391]

For these queries, we need to read one posting list with NSW records from the traditional index for some query lemma. This lemma is designated as the ``main'' lemma of the query.
If there is a frequently used lemma in the query, then we can use two-component key indexes, like (physician, history), for other lemmas. 
If the query consists of only stop and ordinary lemmas, then we use posting lists from the traditional index for the remaining ordinary lemmas but without reading the NSW records. The details are described in \cite{Veretennikov:IntelliSys:2018}.

Average query times:
$Idx0$: 50 s, $Idx5$: 1.9 s, $Idx7$: 2 s, $Idx9$: 2.14 s.

Average data read sizes per query:

$Idx0$: 1.3 GB, $Idx5$: 64.3 MB, $Idx7$: 68.7 MB, $Idx9$: 76 MB.

Average numbers of postings per query:

$Idx0$: 324.7 million, $Idx5$: 3.7 million, $Idx7$: 3.4 million, $Idx9$: 3.3 million.

The number of postings for $Idx7$ is less than that for $Idx5$ because fewer queries were divided into parts. Additionally, the sizes of the NSW records are different for $Idx5$, $Idx7$ and $Idx9$.
We improved the query processing time by a factor of 25.6 with $Idx5$, by a factor of 24.7 with $Idx7$, and by a factor of 23.3 with $Idx9$ in comparison with $Idx0$.

\subsection{Q3. Only frequently used lemmas}

Every query in the subset contains only frequently used lemmas. There are 79 queries in this subset.
Examples include the following:
california mountain pass.

With $FL$-numbers, we have the following query:

[california: 518] [mountain: 704] [pass: 528]

Two-component $(w, v)$ key indexes are used. For example, we can use the (california, pass) and (*pass, mountain) two-component key indexes. 

Average query times:
$Idx0$: 2.5 s, $Idx5$: 0.32 s, $Idx7$: 0.33 s, $Idx9$: 0.34 s.

Average data read sizes per query:

$Idx0$: 53.3 MB, $Idx5$: 4.69 MB, $Idx7$: 4.79 MB, $Idx9$: 4.93 MB.

Average numbers of postings per query:

$Idx0$: 9.4 million, $Idx5$: 0.59 million, $Idx7$: 0.6 million, $Idx9$: 0.61 million.

We improved the query time by a factor of 7.83 with $Idx5$, by a factor of 7.58 with $Idx7$, and by a factor of 7.44 with $Idx9$ in comparison with $Idx0$.
 
\subsection{Q4. Frequently used lemmas and ordinary lemmas}

Every query in the subset contains frequently used lemmas and ordinary lemmas. There are 1 388 queries in this subset.

Examples include the following:
Scalable Vector Graphics.

With $FL$-numbers, we have the following query:

[scalable: $\sim$] [vector: 2953] [graphics: 921]

Each query contains a frequently used lemma; therefore, two-component key indexes can be used. For example, we can use the (graphics, scalable) and (graphics, vector) two-component key indexes. 

Average query times:
$Idx0$: 2.2 s, $Idx5$: 0.36 s, $Idx7$: 0.34 s, $Idx9$: 0.32 s. 

Average data read sizes per query:

$Idx0$: 41.7 MB, $Idx5$: 1.5 MB, $Idx7$: 1.6 MB, $Idx9$: 1.8 MB.

Average numbers of postings per query:

$Idx0$: 8.1 million, $Idx5$: 0.17 million, $Idx7$: 0.17 million, $Idx9$: 0.17 million.

We improved the query time by a factor of 6 with $Idx5$, by a factor of 6.4 with $Idx7$, and by a factor of 6.8 with $Idx9$ in comparison with $Idx0$.

\subsection{Q5. Only ordinary lemmas}

Every query in the subset contains only ordinary lemmas. There are 1 835 queries in this subset.
Examples include the following:
Undersea Fiber Optic Cable.

With $FL$-numbers, we have the following query:

[undersea: 15873] [fiber: 3127] [optic: 2986] [cable: 2771]

We use the traditional index only for these queries. We do not need to read the NSW records because they are stored in separated streams of data. 

Average query times:
$Idx0$: 0.789 s, $Idx5$: 0.819 s, $Idx7$: 0.8 s, $Idx9$: 0.81 s.

Average data read sizes per query:

$Idx0$: 14.2 MB, $Idx5$: 15.6 MB, $Idx7$: 15.7 MB, $Idx9$: 15.8 MB.

Average numbers of postings per query:

$Idx0$: 2.89 million, $Idx5$: 2.89 million, $Idx7$: 2.89 million, $Idx9$: 2.89 million.

The query processing time for these queries does not need any improvement.

\subsection{Results for entire query set}

Average query times:
$Idx0$: 34.9 s, $Idx5$: 1.51 s, $Idx7$: 1.57 s, $Idx9$: 1.66 s.

Average data read sizes per query:

$Idx0$: 0.93 GB, $Idx5$: 46.4 MB, $Idx7$: 49.5 MB, $Idx9$: 54.5 MB.

Average numbers of postings per query:

$Idx0$: 225.7 million, $Idx5$: 3.08 million, $Idx7$: 2.85 million, $Idx9$: 2.76 million.

We improved the query time by a factor of 23.1 with $Idx5$, by a factor of 22.4 with $Idx7$, and by a factor of 21 with $Idx9$ in comparison with $Idx0$.

\subsection{Analysis of the results}

The average time that is needed for the search for $Idx5$, $Idx7$, and $Idx9$, is approximately 1-2 sec. for every query type.

For $Idx0$, we need 2-2.5 sec. for the search if the query consists only of ordinary or frequently used lemmas. For the queries that contain any stop lemma (Q1, Q2), $Idx0$ requires approximately 50 sec. for the search on average. 

This means that the search in $Idx5$, $Idx7$, and $Idx9$ is stable from the performance point of view. However, for $Idx0$, the search system has a performance problem if the query contains any high-frequency occurring lemma. The difference in performance between $Idx0$ and multi-component key indexes $Idx5$, $Idx7$ and $Idx9$ is larger in cases when more high-frequency occurring lemmas occur in the queries.

In \cite{Veretennikov:DAMDID:2018}, we improved the average query execution time by up to 130 times for queries that consist of words occurring with high-frequency. For TREC GOV2, we improved the average query execution time by up to 62.7 times for these queries (i.e., the query set Q1). Perhaps the difference in the ``improvement factor'' is dependent on the average document text size, which is approximately 7 KB for TREG GOV2 and 384 KB for the text collection that was used in \cite{Veretennikov:DAMDID:2018}. The other factor is the length of the query. However, an improvement of 62.7 times in the average query execution time also seems good.

Let us analyze this in more detail.
Let us consider the following example. We have two documents, $D0$ and $D1$.
Let us consider a word $w$. For $w$, we have the following posting list in the traditional index:
$(0,1), (0,5), (0,7), (1,2), (1,5).$

We use two very common encoding schemes.
The idea of the first scheme as follows. We group the records that are related to a specific document.
We convert the original posting list as follows:
$(0, (1,5,7)), (1, (2,5)).$
For other kinds of indexes, we have the same.

Therefore, we store in the index the $ID$ of the document and then the list of word's positions.
The second scheme is a delta-encoding scheme. Consequently, we have the following.
$(0, (1,4,2)), (1, (2,3)).$
For example, consider $(1,5,7)$. Instead of 5, we store $4 = 5-1$, where $1$ is the previous value in the list.

The first scheme is much more effective for text collections that consist of large documents.
This explains the difference in ``improvement factor'' in the experiments with two aforementioned collections.
However, with our method, we can use any other encoding scheme and any other inverted index organization.

In addition, the ``improvement factor'' can depend on the structure of the query set.
To analyze this question, we performed additional experiment.
We formed another query set, using the method from \cite{Veretennikov:DAMDID:2018}.
Consequently, we selected a document from the TREC GOV2 collection. 
We used the content of the document to produce a set of 3500 queries.
We performed our experiments again, using this query set.
The ``improvement factor'' was similar to the foregoing results that we have for TREC GOV2.

The next question is, how can we select the values of our parameters?
The value of $MaxDistance$ affects relevance and should be determined 
according to the selected relevance function \cite{Veretennikov:IntelliSys:2018}. 
The results of experiments allows us to predict how the change of the $MaxDistance$ affects the search time and index size.
Let us discuss now, how the values of $SWCount$ and $FUCount$ affect these metrics.

The value of $SWCount$ is more important than the value of $FUCount$, because it affects Q1 and Q2 queries, which are most complex from the performance point of view.
Let us consider $SWCount$ now.
Let us consider Q2 query subset.
We use Zipf's law \cite{Zipf:1932} as a representation of our word occurrence distribution.
According to this, the second most frequently occurring lemma, occurs half as often as the first.
The third most frequently occurring lemma, occurs 1/3 as often as the first, and so on.
Let $V$ be the number of occurrences of the most frequently occurring lemma.
Therefore, our lemmas have the following number of occurrences: $V, V/2, V/3, ...$

Let us consider a query $q = (q_1, q_2, ..., q_n)$ from Q2.
Here, $q_i$ is the number of corresponding lemma in the $FL$-list.
The query contains one or several stop lemmas and some other lemmas that may be frequently used or ordinary.
To evaluate the query using traditional index, we need to read the following number of postings from the index:
$\sum\limits_{i=1}^n V/q_i$.
Let, without loss of generality, $q_1$ and $q_2$ be stop lemmas and $q_n$ is the main lemma of the query.
With the use of NSW records, we need to read the following number of postings.
$\sum\limits_{i=3}^n V/q_i + (V/q_n) \times (NSWFactor-1)$.
That means, we do not need to read the postings list for $q_1$ and $q_2$. However, for $q_n$ we need to read
its posting list with $NSW$ records. 

The size of the posting $(ID, P, NSW)$ in bytes is up to $NSWFactor = 4.5$ times large than the size of $(ID, P)$. The experiments with $MaxDistance=5$ show this.
Therefore, we can calculate the "planned performance gain" as follows:
$PPG(q) = \left( \sum\limits_{i=1}^n 1/q_i \right) / \left(\sum\limits_{i=3}^n 1/q_i + (1/q_n) \times (NSWFactor - 1)\right).$

If we have a query set $Q$, then we can calculate the average planned performance gain, $APPG(Q) = \frac{1}{|Q|} \sum\limits_{q \in Q} PPG(q)$.
Let now use only Q2 queries for estimations. If a query $q$ is not a Q2 query, then let $PPG(q)$ be 1.
Let $APPG(Q, SWCount)$ be the average planned performance gain that is calculated for a specific value of $SWCount$.

For the query set that we use in this paper, i.e., 10 665 queries, we have the following:
$APPG(Q, 100) = 43$, $APPG(Q, 500) = 105$, $APPG(Q, 1000) = 153.$
However, this model do not take into account $(w, v)$ and $(f,s,t)$ indexes. 
In future, we plan to develop more precise models.
However, from this model, we can predict that $SWCount$ should not be increased.

The foregoing results need a more detailed examination. Let us consider a search query. The query consists of some set of lemmas. Let $Min$-$FL$-number be the minimum $FL$-number among all lemmas of the query. A lower $FL$-number corresponds to a more frequently occurring lemma. If the $Min$-$FL$-number of a query is a small number, then the query can induce performance problems because the query contains some high-frequency occurring lemma.

Then, we divide the entire query set into subsets based on the $Min$-$FL$-number of queries. We select 100 as the division step. In the first subset, we include all queries with $Min$-$FL$-numbers from 0 to 99; in the second subset, we include all queries with $Min$-$FL$-numbers from 100 to 199; and so on. In the following diagrams, we consider the first 21 subsets.
 
 \begin{figure}[h]
 \setlength{\abovecaptionskip}{1pt}
 \setlength{\belowcaptionskip}{-10pt}
  \centering
  \includegraphics[width=0.7\linewidth]{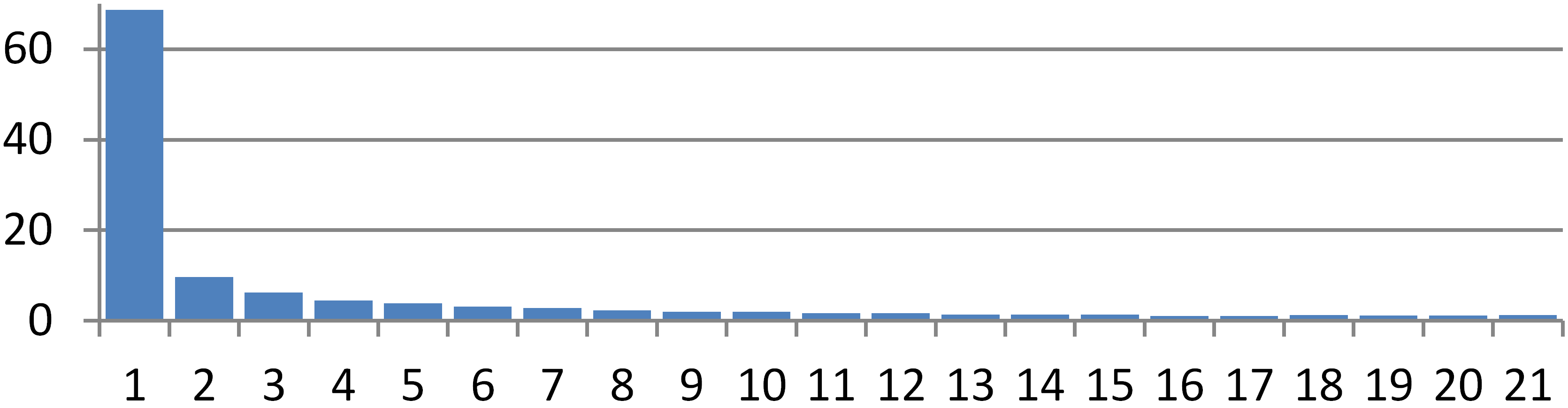}
  \caption{The average query execution times for $Idx0$ (seconds); the query set is divided based on the $Min$-$FL$-number with a step of 100.}
  \label{VeretennikovA-image-step-Idx0-Search}
\end{figure}

In Fig.~\ref{VeretennikovA-image-step-Idx0-Search}, the average query execution time for $Idx0$ in seconds is displayed for every subset. The first bar is significantly larger than the other bars, showing that the first subset induces some performance problems. The first 8 subsets have an average query execution time of more than two seconds.

\begin{figure}[h]
\setlength{\abovecaptionskip}{1pt}
\setlength{\belowcaptionskip}{-10pt}
  \centering
  \includegraphics[width=0.7\linewidth]{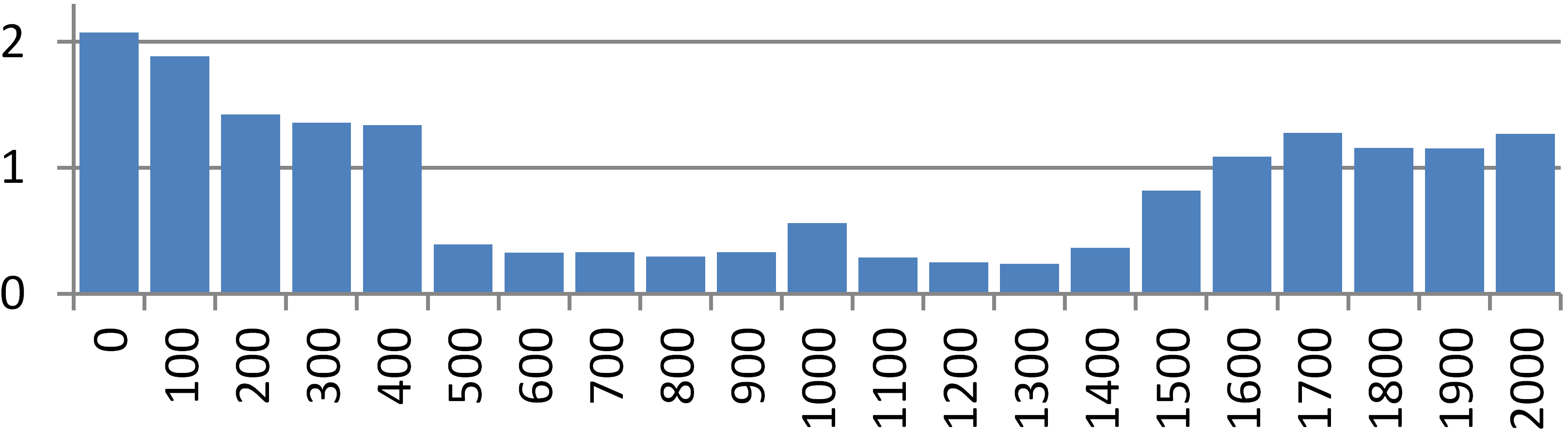}
  \caption{The average query execution times for $Idx5$ (seconds); the query set is divided based on the $Min$-$FL$-number with a step of 100.}
  \label{VeretennikovA-image-step-Idx5-Search}
\end{figure}
 
In Fig.~\ref{VeretennikovA-image-step-Idx5-Search}, the average query execution time for $Idx5$ in seconds is displayed for every subset. For the first subset, the average query execution time is about two seconds. For each following subset, the average query execution time is less than two seconds.

\begin{figure}[h]
\setlength{\abovecaptionskip}{1pt}
\setlength{\belowcaptionskip}{-10pt}
  \centering
  \includegraphics[width=0.7\linewidth]{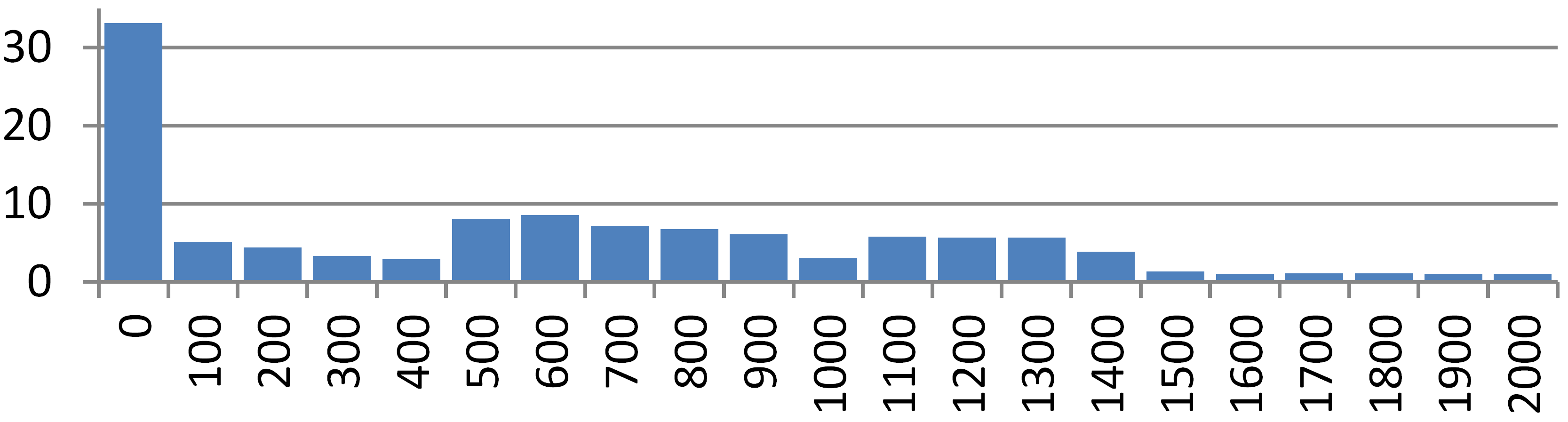}
  \caption{The improvement factor for $Idx5$ in comparison with $Idx0$ (times); the query set is divided based on the $Min$-$FL$-number with a step of 100.}
  \label{VeretennikovA-image-step-Idx5-Search-Factor}
\end{figure}

In Fig.~\ref{VeretennikovA-image-step-Idx5-Search-Factor}, we show the improvement factor for $Idx5$ in comparison with $Idx0$. The first bar value is 33, which means that the average query execution time for the first subset is improved by a factor of 33. We also see how the performance is changed when the $Min$-$FL$-number of a query crosses the $SWCount$ of 500.

By this diagram, we can propose that the value of $SWCount$ can be lowered to 100.
Consequently, we created another index $Idx5/SW100$, with $SWCount = 100$ and $FUCount = 1450$.
  
 \begin{figure}[h]
 \setlength{\abovecaptionskip}{1pt}
 \setlength{\belowcaptionskip}{-10pt}
  \centering
  \includegraphics[width=0.7\linewidth]{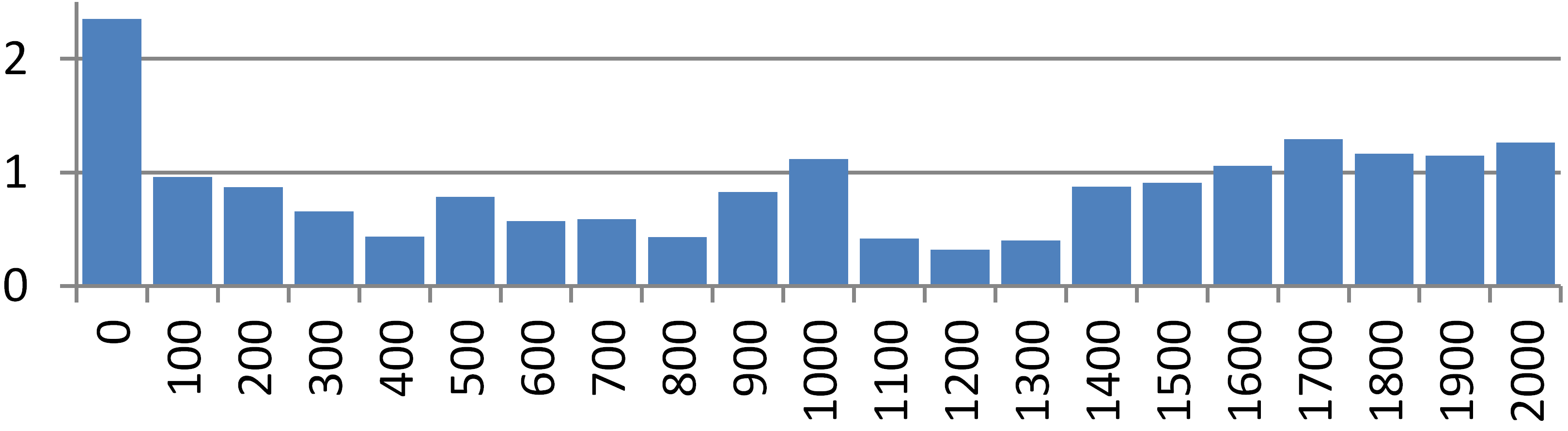}
  \caption{The average query execution times for $Idx5/SW100$ (seconds); the query set is divided based on the $Min$-$FL$-number with a step of 100.}
  \label{VeretennikovA-image-step-Idx5-100-Search}
\end{figure}

In Fig.~\ref{VeretennikovA-image-step-Idx5-100-Search}, the average query execution time for $Idx5$/SW100 in seconds is displayed for every subset. For each subset, the average query execution time is less than two seconds. The results look more promising than those for $Idx5$. 

However, let us consider Q1 queries when the value of $SWCount$ is 500, which are the queries that consist only of lemmas with $FL$-number $<$ 500. There are 119 queries in this subset, as discussed above. It was established that for this subset, the average query search time with $Idx5$ is 0.8 sec. but with $Idx5/SW100$, it is 6.7 sec., which is significantly larger. When this subset is evaluated with $Idx5$, three-component key indexes are used.
  
 \begin{figure}[h]
 \setlength{\abovecaptionskip}{1pt}
 \setlength{\belowcaptionskip}{-10pt}
 \setlength{\textfloatsep}{1pt} 
  \centering
  \includegraphics[width=0.7\linewidth]{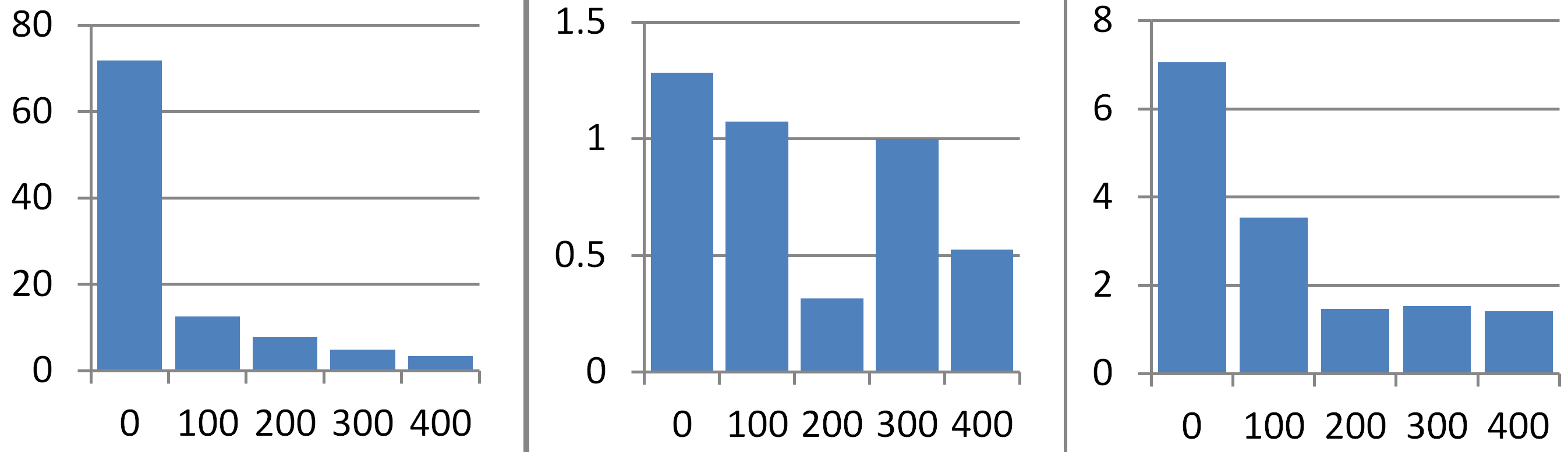}
  \caption{The average query execution times for $Idx0$, $Idx5$, and $Idx5/SW100$ (seconds); the query set Q1 is divided based on the $Min$-$FL$-number with a step of 100.}
  \label{VeretennikovA-image-step-Idx0-Idx5-Idx5-100-Q0-Search}
\end{figure} 

Let us divide query set Q1 into five subsets based on the $Min$-$FL$-number value of the concrete query. In Fig.~\ref{VeretennikovA-image-step-Idx0-Idx5-Idx5-100-Q0-Search}, the average query execution time for $Idx0$, $Idx5$ and $Idx5$/SW100 in seconds is displayed for every subset. We see that the search in $Idx5$/SW100 works significantly more slowly than that in $Idx5$.

Therefore, we come to the following conclusions for different kinds of queries.
Consequently, we propose the new index schema. 

\vspace{-\topsep}
\vspace{1pt}
\begin{enumerate}
 \setlength{\itemsep}{1pt}
  \setlength{\parskip}{1pt}
  \setlength{\parsep}{1pt}
  \item[T1]  
	A query consists of extreme high-frequency occurring lemmas and
    high-frequency occurring lemmas. This means that all lemmas of the query are extreme high-frequency occurring or high-frequently occurring. E.g., for every lemma of the query, $FL$-number $<$ 500.
  
    The $(f, s, t)$ indexes work better for these queries than do other types of indexes, as shown by Fig.~\ref{VeretennikovA-image-step-Idx0-Idx5-Idx5-100-Q0-Search}. The search in $Idx5/SW100$ is significantly slower than that in $Idx5$. When the search is performed using $Idx5$, the $(f, s, t)$ indexes are used for every query in T1. When the search is performed using $Idx5/SW100$, the $(f, s, t)$ indexes are used only if $FL(f) < 100$, $FL(s) < 100$, and $FL(t) < 100$.
\item[T2]
A query contains a low-frequency occurring lemma (ordinary or frequently used). Additionally, the query may contain high-frequency occurring lemmas, but for every query lemma, $FL$-number $\geq$ 100.

The $(w, v)$ indexes allow one to achieve a better performance improvement, as shown in Fig.~\ref{VeretennikovA-image-step-Idx5-100-Search}. For subsets starting from $FL$-number = 100, no $(f, s, t)$ indexes or NSW records are used when we do the search using $Idx5/SW100$. Here, $Idx5/SW100$ is better than $Idx5$.
\item[T3]
A query contains a low-frequency occurring lemma (ordinary or frequently used) and some extreme high-frequency occurring lemmas with $FL$-number $<$ 100.

The $(w, v)$ indexes and NSW records are required. The first bar in Figures \ref{VeretennikovA-image-step-Idx0-Search}, \ref{VeretennikovA-image-step-Idx5-Search} and \ref{VeretennikovA-image-step-Idx5-100-Search} supports this, and our previous experiments in \cite{Veretennikov:DAMDID:2018} confirm it.
\end{enumerate} 

\subsection{The new index schema}
The original schema can be represented by the following rules:

\vspace{-\topsep}
\vspace{1pt}
\begin{enumerate}
    \setlength{\itemsep}{1pt}
    \setlength{\parskip}{1pt}
    \setlength{\parsep}{1pt}
\item[1)]	$(f, s, t)$ indexes, where 
$FL(f), FL(s), FL(t) < SWCount$.

\item[2)]	$(w, v)$ indexes, where

\centerline{$SWCount \leq FL(w) < SWCount + FUCount, \textnormal{ and }
SWCount \leq FL(v).$}

\item[3)]	Traditional indexes $(x)$ with NSW records, $SWCount \leq FL(x)$; the NSW records contain information about all lemmas $y$ with the condition $FL(y) < SWCount$ that occur near lemma $x$ in the text.
\end{enumerate}

For the new schema, we use the following parameters with example values.

$EHFCount = 100$ -- for extreme high-frequency occurring lemmas.

$HFCount = 400$ -- for high-frequency occurring lemmas.

$FUCount = 1050$ -- for frequently used lemmas.

We propose using the following indexes.

\vspace{-\topsep}
\vspace{1pt}
\begin{enumerate}
    \setlength{\itemsep}{1pt}
    \setlength{\parskip}{1pt}
    \setlength{\parsep}{1pt}
\item[1)]	$(f, s, t)$ indexes that can be used for T1 queries, where 

\centerline{$FL(f), FL(s), FL(t) < EHFCount+ HFCount = 500, $}

\item[2)]	$(w, v)$ indexes that can be used for T2 and T3 queries, where 

\centerline{$100 = EHFCount \leq FL(w) < 
EHFCount+HFCount+FUCount = 1450, $}

\centerline{$100 = EHFCount \leq FL(v).$}

\item[3)]	Traditional indexes $(x)$ with NSW records, $100 = EHFCount \leq FL(x)$; the NSW records contain information about all lemmas $y$ with the condition $FL(y) < EHFCount  = 100$ that occur near lemma $x$ in the text.
These indexes can be used for T3 queries.
\end{enumerate}

The concrete values of the parameters are provided only for example and can be different for different languages and text collections.

\section{Conclusions and future work}
In this paper, we investigated how multi-component key indexes help to improve search performance. We used well-known GOV2 text collection. We proposed a method of analyzing the search performance by considering different types of queries. By following this method, we found that the performance can be improved further and proposed a new index schema.

We analyzed how the value of $MaxDistance$ affects the search performance. With an increase in the value of $MaxDistance$ from 5 to 9, the average search time using multi-component key indexes was increased from 1.51 sec. to 1.66 sec. Therefore, the value of $MaxDistance$ can be increased even further, and the main limitations here are the disk space and the time of indexing.
Our multi-component indexes are relatively large for large values of $MaxDistance$. However, large hard disk drives are now available. In many cases, it would be preferable to spend several TB of disk space but to increase the search speed by a factor of 20 times or more.

We found that multi-component key indexes work significantly better on text collections with large documents (e.g., documents with sizes of approximately several hundred KB or more) than on text collection that consists of small documents; thus, an improvement factor of 20 can be considered as a minimum improvement factor.
In the future, it is important to consider how different compression technologies can reduce the index total size, which will increase the search speed even more.

The proposed indexes with multi-component keys have one limitation. If we have a document that contains queried words and the distance between these words is greater than $MaxDistance$, then this document can be absent in the search results. This is usually not a problem if the average document size in the text collection is relatively large, e.g., several hundreds of kilobytes. In this case, after the proximity search with multi-component key indexes, we can run a search without distance. When the former requires the word-level index, the latter needs only the document-level index and works significantly faster. 

Second, in the majority of modern relevance models, it is defined that the weight of the document is inversely proportional to the square of the distance between queried words in the document \cite{Yan:2010:ETP:1871437.1871593}. With a relatively large value of $MaxDistance$, we can be sure that all relevant documents will occur in the search results. When the first consideration for GOV2 collection is under question, because the documents are small, the second is still valid. 

\bibliographystyle{splncs04}

\end{document}